\documentclass{aa}
\usepackage{graphics,epsfig,times}

\usepackage{natbib}
\bibpunct{(}{)}{;}{a}{}{,}

\newcommand{\ergcm}[1]{$\times 10^{#1}$ erg cm$^{-2}$ s$^{-1}$}

\newcommand{\hcm}[1]{$\times 10^{#1}$ cm$^{-2}$}

\newcommand{\expo}[1]{$\times 10^{#1}$}
\newcommand{\oexpo}[1]{$10^{#1}$}
\newcommand{\nh}{N$_{\rm H}$}

\newcommand{\ltsima}{$\buildrel < \over \sim$}
\newcommand{\lsim}{\lower.5ex\hbox{\ltsima}}
\newcommand{\gtsima}{$\buildrel > \over \sim$}
\newcommand{\gsim}{\lower.5ex\hbox{\gtsima}}

\newcommand{\xnorth}{\hbox{\object{XMMU\,J005517.9$-$723853}}}
\newcommand{\xsouth}{\hbox{\object{XMMU\,J005455.4$-$724512}}}
\newcommand{\xna}{\hbox{J0055$-$7238}}
\newcommand{\xsa}{\hbox{J0054$-$7245}}
\newcommand{\xte}{\hbox{\object{XTE\,J0055$-$727}}}
\begin{document}
 
\title{Two long-period X-ray pulsars detected in the SMC field around \xte
       \thanks{Based on observations with 
               XMM-Newton, an ESA Science Mission with instruments and contributions 
               directly funded by ESA Member states and the USA (NASA)}}
 
\author{F.~Haberl\inst{1} \and W.~Pietsch\inst{1} \and 
        N.~Schartel\inst{2} and P.~Rodriguez\inst{2} and 
        R.H.D. Corbet\inst{3,4}}

\titlerunning{Two long-period X-ray pulsars detected in the SMC field around \xte}
\authorrunning{Haberl et al.}
 
\offprints{F. Haberl, \email{fwh@mpe.mpg.de}}
 
\institute{Max-Planck-Institut f\"ur extraterrestrische Physik,
           Giessenbachstra{\ss}e, 85748 Garching, Germany
	   \and
	   XMM-Newton Science Operations Centre, ESA, Villafranca del Castillo, 
	   PO Box 50727, 28080 Madrid, Spain
	   \and
	   Laboratory for High Energy Astrophysics, Code 662, NASA Goddard Space Flight Center, 
	   Greenbelt, MD 20771, USA
	   \and
	   Universities Space Research Association
	   }
 
\date{Received 17 February 2004 / Accepted 30 April 2004}
 
\abstract{
An XMM-Newton target of opportunity observation of the field around the transient 18.37 s 
pulsar \xte\ in the Small Magellanic Cloud (SMC) revealed two bright, long-period X-ray 
pulsars in the EPIC data. A new pulsar, \xnorth, with a pulse period of 701.7 $\pm$ 0.8 s 
was discovered and 500.0 $\pm$ 0.2 s pulsations were detected from \xsouth\ 
(= CXOU\,J005455.6$-$724510), confirming the period found in Chandra data. We derive 
X-ray positions of RA = 00$^{\rm h}$54$^{\rm m}$55\fs88, Dec = --72\degr45\arcmin10\farcs5 and 
RA = 00$^{\rm h}$55$^{\rm m}$18\fs44, Dec = --72\degr38\arcmin51\farcs8 (J2000.0) 
with an uncertainty of 0\farcs2 utilizing optical identification with OGLE stars.
For both objects, the optical brightness and colours and the X-ray spectra are consistent 
with Be/X-ray binary systems in the SMC.

}
 
\maketitle
 
\section{Introduction}

The Small Magellanic Cloud harbours a large number of high mass 
X-ray binary (HMXB) systems, more than are known in the Large Magellanic Cloud and the 
Milky Way \citep[see compilations by ][]{2000A&A...359..573H,2003PASJ...55..161Y}
despite the much smaller mass of the SMC. The catalogue of 
\citet{2004A&A...414..667H} comprises 65 HMXBs and candidates in the SMC with at least 37 
showing X-ray pulsations which indicate the spin period of the neutron star in orbit
around a high mass early type star. The number of X-ray 
pulsars in the SMC has meanwhile further grown by five: For three of the known candidate 
HMXBs pulsations of 
202 s, 500 s and 138 s were detected, 
an X-ray source known from ROSAT was discovered as 34.1 s pulsar 
and 18.37 s pulsations were found in RXTE observations of the SMC. 
In addition accurate Chandra positions enabled the location of
two RXTE-discovered pulsars with 82.4 s and 
7.78 s period, the latter identified with SMC\,X-3
\citep[see][ and references therein]{2004astroph0402053C}.
 
The non-imaging instruments on RXTE allowed the 18.37 s transient pulsar 
\xte\ to be located to an accuracy of only 0.1\degr$\times$0.06\degr. Hence, we proposed 
an XMM-Newton target of opportunity observation which was performed on December 18, 2003.
We could not detect 18.37 s pulsations from any of the weaker sources seen in 
the EPIC images due to insufficient counting statistics and were therefore not able 
to identify the target. However, the two brightest X-ray sources were found to 
exhibit pulsations with longer periods \citep{2004ATel..219....1H}. 
\xnorth\ (hereafter \xna) is a new pulsar and \xsouth\ (hereafter \xsa) is identified 
with CXOU\,J005455.6$-$724510 by position and X-ray period of $\sim$500 s.
Here we present the results of a temporal and spectral analysis of the X-ray data of these two SMC
pulsars and propose optical counterparts with properties as expected for Be/X-ray binary systems. 

\section{XMM-Newton observation and analysis}

The XMM-Newton observation of the field around the transient pulsar \xte\ in the 
SMC was performed on December 18, 2003 between 14:32 UT and 19:48 UT.
The EPIC-pn instrument \citep{2001A&A...365L..18S} was operated in large window
imaging mode for a net exposure of 16.1 ks, while EPIC-MOS1 and MOS2  
\citep{2001A&A...365L..27T} were operated in full frame imaging mode 
for 14.4 ks and 18.5 ks, respectively.
For optical light blocking the medium filter was used in all cameras.
The data were processed using the XMM-Newton analysis package SAS version 5.4.1 to 
produce the photon event files and binned data products such as images, spectra and 
light curves. Events for spectral and temporal analysis of the two pulsars were 
extracted from circular regions (radius 30\arcsec) around the source positions
and from nearby source-free regions for background spectra.

\subsection{X-ray pulsations}

\begin{figure*}
\resizebox{6.2cm}{!}{\includegraphics[clip=]{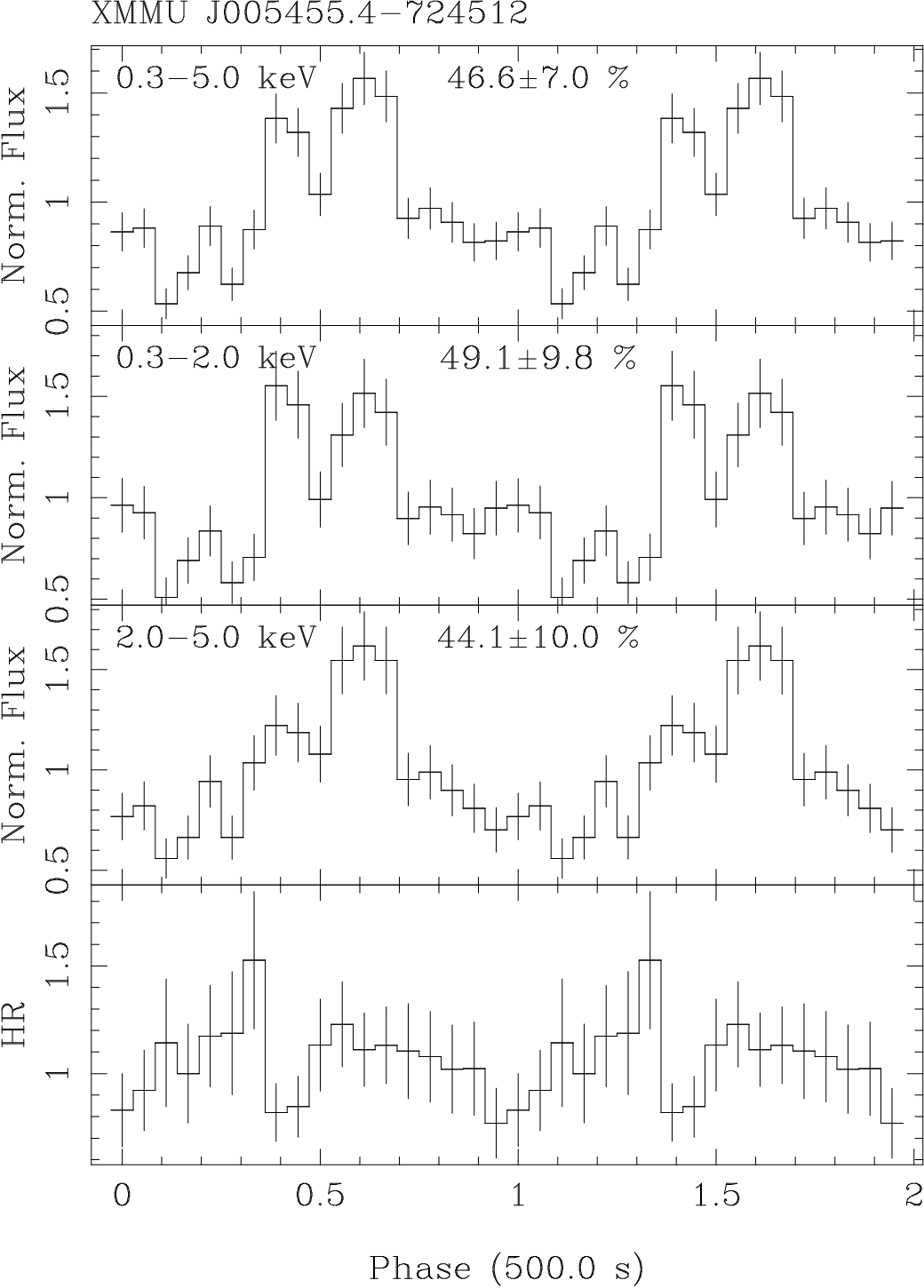}}
\hspace{1mm}
\resizebox{6.2cm}{!}{\includegraphics[clip=]{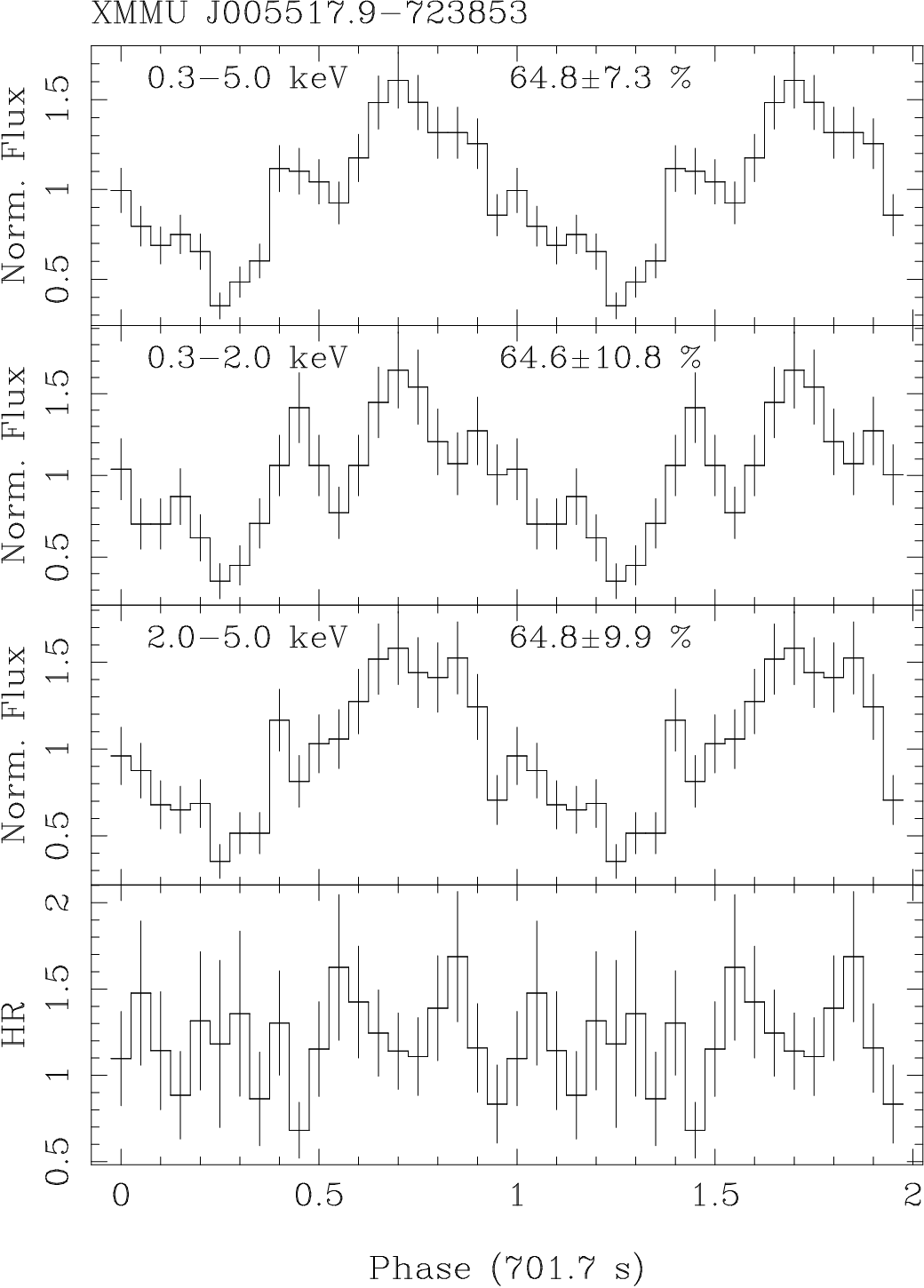}}
  \hfill
\parbox[b]{50mm}{
\caption{EPIC-pn light curves folded at the pulse period in broad, hard and soft 
         energy bands together with the hardness ratio, the ratio of the
         count rates in the hard to the soft band. The pulsed fraction, defined as (mean - minimum)/mean), 
	 in each energy band is given in \%.}}
\label{epic-hardness}
\end{figure*}

We performed a timing analysis of the EPIC data of the two bright sources visible in the images
following the approach of \citet[][and references therein]{2000ApJ...540L..25Z}.
First, we searched for pulsations in the broad band (0.3$-$5.0 keV) EPIC-pn data using the 
Rayleigh Z$^2_1$ technique in the range 0.001-4 Hz. Strong peaks in the probability density 
function were found with Z$^2_1$ of 108.9 and 126.5, which 
correspond to a period detection confidence of 1.0-1.5\expo{-19} and 1.0-2.3\expo{-23} for \xsa\ 
and \xna, respectively (neglecting that the period of \xsa\ was already known). 
Then, periods and 1$\sigma$ errors were determined using the Odds-ratio method based on the Bayesian 
formalism to 701.7$\pm$0.8 s (\xna) and 500.0$\pm$0.2 s (\xsa).
EPIC-pn light curves in different energy bands (0.3-2.0 keV, 2.0-5.0 keV and the total 0.3-5.0 keV) 
were folded with the pulse period and are shown in Fig.~\ref{epic-hardness}. The broad band pulse
profile is highly structured for both objects with some indication for hardness ratio changes 
during individual dips/peaks. Although over the whole pulse period the hardness ratio is formally consistent with 
a constant value, at certain phases the hardness ratio changes over 3-4 phase bins accumulating 
3-4 $\sigma$ deviations from the average (\xsa: phase 0.4; \xna: phases 0.6 and 0.85).

\subsection{X-ray spectra}

\begin{figure*}
\begin{center}
\resizebox{8.7cm}{!}{\includegraphics[clip=,angle=-90]{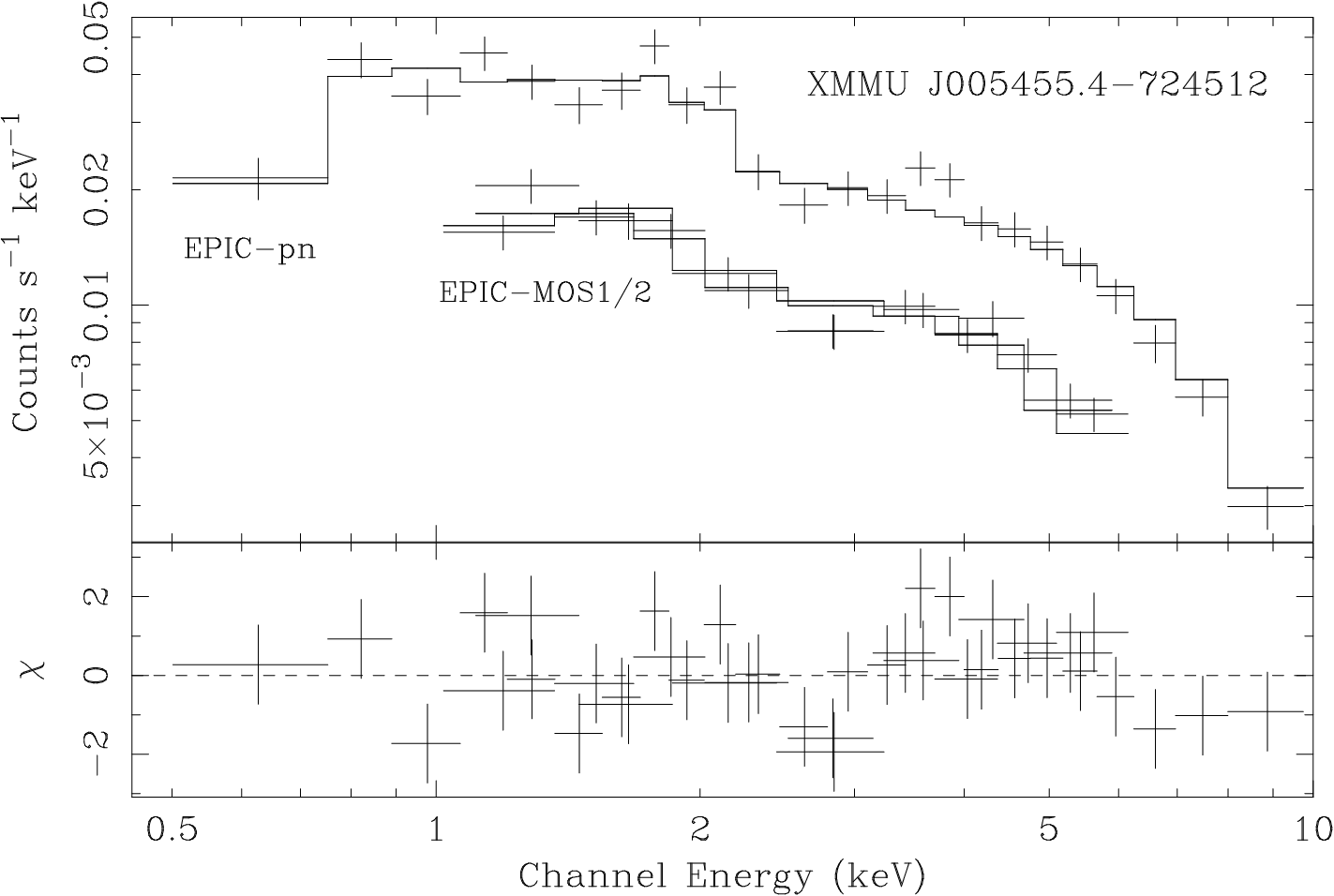}}
\hspace{3mm}
\resizebox{8.7cm}{!}{\includegraphics[clip=,angle=-90]{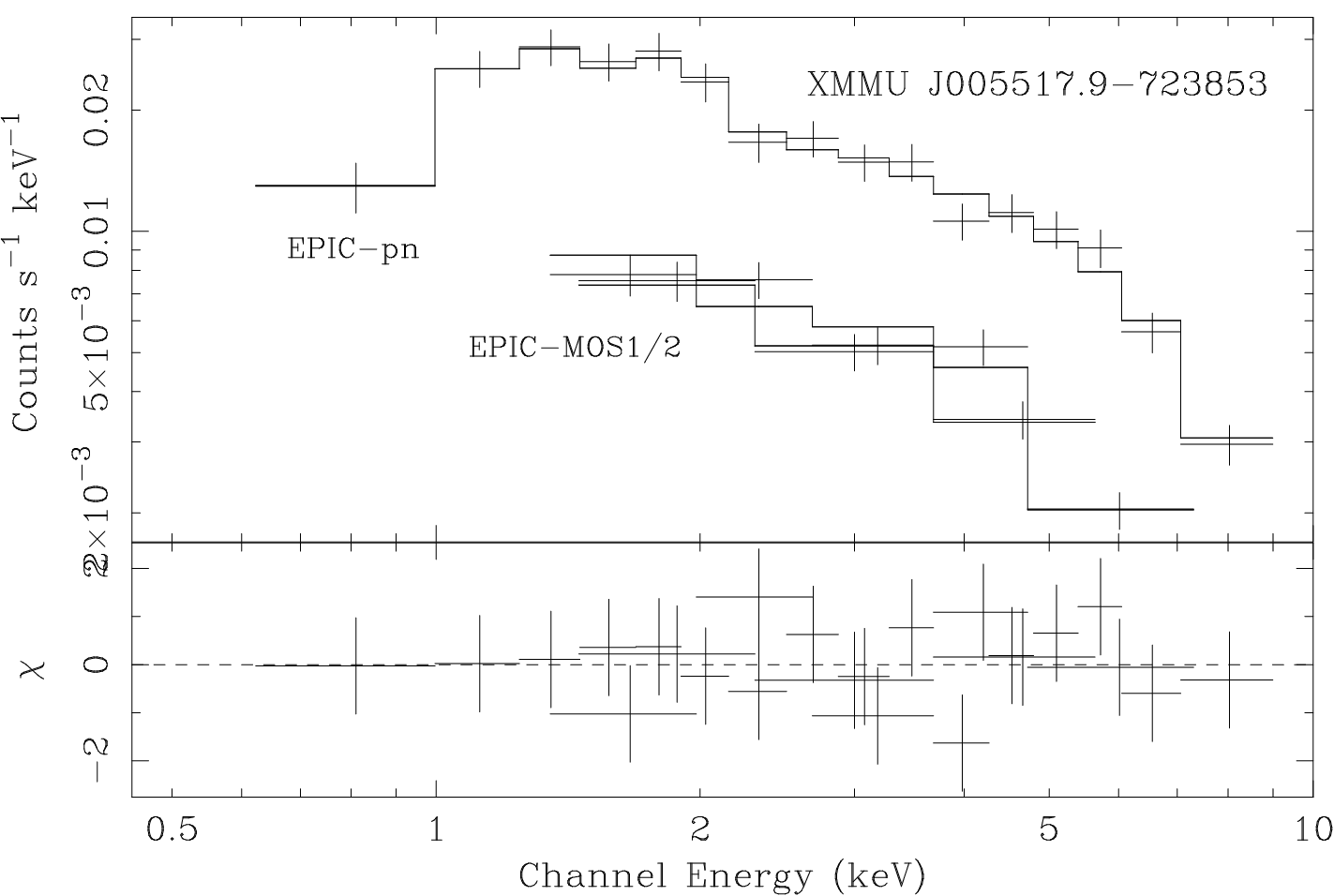}}
\end{center}
\caption{EPIC spectra of the two pulsars. The histograms show the best-fit model comprising a power-law and
a thermal plasma emission component. The total number of counts used in the spectral analysis was 3943 and 2301 for
\xsa\ and \xna, respectively.}
\label{epic-spectra}
\end{figure*}

Pulse phase averaged EPIC-pn and -MOS spectra were extracted using single+double pixel events (pattern 0-4) 
and pattern 0-12, respectively and binned to obtain at least 100 counts per bin. The three spectra for each 
pulsar were fit simultaneously (using XSPEC v11.3) with the same model only allowing a free normalization 
factor between the different instruments. 
Errors were determined for 90\% confidence levels.
A simple power-law (PL) model including photo-electric absorption by matter with solar abundance did not 
reproduce the data well (reduced $\chi^2 > 1.58$, see Table~\ref{fit-spectra}). Then other
models which were used in the past to fit X-ray spectra of HMXBs were applied: adding a soft component 
like black-body (BB) or thermal plasma (MEKAL) emission or a power-law with exponential high-energy cutoff 
(PL*EXP). For both objects the fits with the latter three models are acceptable but do not allow us to
formally differentiate between them. In Table~\ref{fit-spectra} the characteristic parameters are listed
(photon index $\gamma$, temperature kT, high energy cutoff E$_{\rm cut}$ and folding energy E$_{\rm fold}$)
together with observed fluxes and intrinsic source luminosities (only for the models with acceptable fit)
and reduced $\chi^2$ values per degree of freedom (dof). The EPIC spectra together with the PL plus MEKAL 
models (the other acceptable models do not look different) are plotted in Fig.~\ref{epic-spectra}.

\begin{table*}
\caption[]{Spectral fit results.}
\begin{center}
\begin{tabular}{lccccccccc}
\hline\hline\noalign{\smallskip}
\multicolumn{1}{l}{Model$^{(1)}$} &
\multicolumn{1}{c}{\nh} &
\multicolumn{1}{c}{$\gamma$} &
\multicolumn{1}{c}{kT} &
\multicolumn{1}{c}{E$_{\rm cut}$} &
\multicolumn{1}{c}{E$_{\rm fold}$} &
\multicolumn{1}{c}{Flux$^{(2)}$} &
\multicolumn{1}{c}{L$_{\rm x}^{(3)}$} &
\multicolumn{1}{c}{L$_{\rm x}^{\rm i~(4)}$} &
\multicolumn{1}{c}{$\chi^2_{\rm r}$/dof} \\

\multicolumn{1}{l}{} &
\multicolumn{1}{c}{[\oexpo{22}cm$^{-2}$]} &
\multicolumn{1}{c}{} &
\multicolumn{1}{c}{[keV]} &
\multicolumn{1}{c}{[keV]} &
\multicolumn{1}{c}{[keV]} &
\multicolumn{1}{c}{erg cm$^{-2}$ s$^{-1}$} &
\multicolumn{1}{c}{erg s$^{-1}$} &
\multicolumn{1}{c}{erg s$^{-1}$} &
\multicolumn{1}{c}{} \\

\noalign{\smallskip}\hline\noalign{\smallskip}
\multicolumn{6}{l}{\xsouth} \\
 PL	  & 0.15		   & 1.05	   & -- 		       & --          & --                  & --		    & --	   & --  	  & 2.03/35 \\
 PL+BB    & 0.83$^{+0.15}_{-0.22}$ & 1.05$\pm$0.13 & 0.095$^{+0.015}_{-0.010}$ & --          & --                  & 1.45\expo{-12} & 6.3\expo{35} & 4.9\expo{36} & 1.25/33 \\
 PL+MEKAL & 0.90$^{+0.16}_{-0.26}$ & 1.02$\pm$0.12 & 0.18$^{+0.06}_{-0.04}$    & --          & --                  & 1.46\expo{-12} & 6.3\expo{35} & 5.3\expo{36} & 1.30/33 \\
 PL*EXP   & 0.06$\pm$0.05          & 0.40$\pm$0.14 & --                        & 3.9$\pm$0.9 & 6.5$^{+2.8}_{-1.6}$ & 1.39\expo{-12} & 6.0\expo{35} & 6.2\expo{35} & 1.18/33 \\
\hline\noalign{\smallskip}
\multicolumn{6}{l}{\xnorth} \\
 PL	  & 0.45		   & 0.96	   & -- 		       & --          & --                  & --		    & --	   & --  	  & 1.58/19 \\
 PL+BB    & 1.3$\pm$0.4 	   & 1.31$\pm$0.20 & 0.10$\pm$0.02	       & --          & --                  & 9.22\expo{-13} & 4.0\expo{35} & 6.5\expo{36} & 0.81/17 \\
 PL+MEKAL & 1.7$\pm$0.4 	   & 1.42$\pm$0.19 & 0.13$^{+0.06}_{-0.03}$    & --          & --                  & 9.11\expo{-13} & 3.9\expo{35} & 1.1\expo{38} & 0.72/17 \\
 PL*EXP   & 0.19$\pm$0.15          & 0.26$^{+0.4}_{-0.9}$ & --                 & 2.6$\pm$1.6 & 5.2$^{+4.4}_{-2.7}$ & 8.98\expo{-13} & 3.9\expo{35} & 4.1\expo{35} & 0.84/17 \\
\noalign{\smallskip}\hline
\end{tabular}
\end{center}
$^{(1)}$ For definition of spectral models see text. $^{(2)}$ Observed 0.5-10.0 keV flux.
$^{(3)}$ X-ray 0.5-10.0 keV luminosity (including absorption). \\ 
$^{(4)}$ Source intrinsic X-ray luminosity in the 0.1-10.0 keV band (corrected for absorption) 
for a distance of 60 kpc.
\label{fit-spectra}
\end{table*}

\subsection{Source identifications}

X-ray source positions were obtained from a combined analysis of the EPIC images 
and are given by \citet{2004ATel..219....1H}. The uncertainties are
dominated by the $\sim$3\arcsec\ systematic bore-sight error (90\% confidence\footnote{
See the statistical analysis of the 1XMM catalogue prepared by the XMM-Newton Survey 
Science Centre Consortium ({\tt http://xmmssc-www.star.le.ac.uk})}). In the following 
we derive improved X-ray coordinates by reducing this systematic uncertainty. 

In a first step the accurate (0.6\arcsec) Chandra position of \xsa\ 
(= CXOU\,J005455.6$-$724510) enables us to optically identify this 500 s pulsar. \xsa\ was 
also detected by Chandra when it was found to be pulsating with a period of 503.5$\pm$6.7 s 
\citep{2004ATel..217....1E}. This clearly establishes that both sources are identical.
The position is also within the error circles of AX\,J0054.8$-$7244 \citep{2003PASJ...55..161Y}
and of RX\,J0054.9$-$7245 which was proposed as HMXB
candidate by \citet{2000A&A...359..573H} due to the presence of a close emission line object
\citep[MA93\#809][]{1993A&AS..102..451M}. 
Using the finding chart provided by these authors allows the identification of 
MA93\#809 with a star 
covered by the SMC UBVR CCD survey of \citet{2002ApJS..141...81M} and also listed in 
the OGLE BVI photometry catalogue \citep{1998AcA....48..147U}. Optical positions and 
available magnitudes are summarized in Table~\ref{tab-ids}. 
The presence of a $\sim$15 mag star inside the error circle of the Chandra position 
clearly establishes it as the optical counterpart of the X-ray source (detected by 
Chandra and XMM).

The 701 s pulsar \xna\ is located within the 9\arcsec\ error radius of the ROSAT source
RX\,J0055.2$-$7238 \citep{2000A&AS..142...41H}. 
The ROSAT PSPC hardness ratios indicated a hard source but were tainted with large errors.
No archival Chandra observation covers the source. 
After applying the boresight correction found for \xsa\ above (with a remaining uncertainty
of $\sim$0.3\arcsec) to \xna, also for this source a bright optical counterpart is found in 
the catalogues used above with properties given in Table~\ref{tab-ids}.
The fact that in both cases objects with brightness and colours as expected for Be star companions
are found in the sub-arcsecond error circles strongly supports their correct identification.
In a final step we use both objects to determine the most precise source positions (additional
much fainter X-ray sources in the field have larger statistical errors and do not allow 
further improvement). 
For both pulsars the originally EPIC derived X-ray position is systematically shifted 
with respect to the optical position by +0\fs5 in RA and \hbox{+1\farcs4 /+1\farcs7} in Declination. 
After correction we derive final X-ray coordinates of 
RA = 00$^{\rm h}$54$^{\rm m}$55\fs88 and Dec = --72\degr45\arcmin10\farcs5 (J2000.0) for \xsa\ and
RA = 00$^{\rm h}$55$^{\rm m}$18\fs44 and Dec = --72\degr38\arcmin51\farcs8 (J2000.0) for \xna.
The remaining uncertainty (including 0.13\arcsec\ statistical error) is 0.2\arcsec.

The inferred effective temperature and bolometric luminosity of the optical counterparts given by
\citet{2002ApJS..141...81M} suggest spectral types of O9 V for both stars. Assuming this spectral type 
the measured B$-$V indices imply E(B$-$V) of 0.28 and 0.48 and \nh\ = 1.6\hcm{21} and 2.8\hcm{21}
for \xsa\ and \xna, respectively.

\begin{table*}
\caption[]{Optical identifications.}
\begin{center}
\begin{tabular}{lccccccc}
\hline\hline\noalign{\smallskip}
\multicolumn{1}{l}{Source} &
\multicolumn{1}{c}{Catalogue} &
\multicolumn{1}{c}{R.A. and Dec. (2000.0)} &
\multicolumn{1}{c}{Vmag} &
\multicolumn{1}{c}{B$-$V} &
\multicolumn{1}{c}{U$-$B} &
\multicolumn{1}{c}{V$-$R} &
\multicolumn{1}{c}{V$-$I} \\

\noalign{\smallskip}\hline\noalign{\smallskip}
\xsouth & UBVR  & 00$^{\rm h}$54$^{\rm m}$55\fs88 --72\degr45\arcmin10\farcs5 & 14.78 & $-$0.05 & $-$0.90 & $-$0.86 & --   \\
        & OGLE  & 00$^{\rm h}$54$^{\rm m}$55\fs87 --72\degr45\arcmin10\farcs7 & 14.99 & $-$0.02 & --      & --      & 0.20 \\
\hline\noalign{\smallskip}
\xnorth & UBVR  & 00$^{\rm h}$55$^{\rm m}$18\fs47 --72\degr38\arcmin51\farcs6 & 15.87 & $-$0.15 & $-$0.83 & $-$0.94 & --   \\
        & OGLE  & 00$^{\rm h}$55$^{\rm m}$18\fs43 --72\degr38\arcmin51\farcs8 & 16.01 & $-$0.08 & --      & --      & 0.30 \\
\noalign{\smallskip}\hline
\end{tabular}
\end{center}
\label{tab-ids}
\end{table*}

\section{Discussion}

The December 2000 XMM-Newton observation of the SMC region around \xte\ revealed 
two X-ray pulsars with long pulse periods.  
The 500 s pulsar \xsa\ was the brighter of the two and was 
detected by Chandra on July 4, 2002, with pulsations reported by \citet{2004ATel..217....1E}. 
It was detected during three ROSAT observations in May 1993, April 1994 and April/May 1997.
Using the parameters derived from the EPIC spectra and the ROSAT PSPC and HRI spectral
response to estimate expected ROSAT count rates, shows that this pulsar exhibits variations
in flux by about a factor of 3 with intensities lowest in May 1993 
and highest in December 2000. During several other ROSAT observations the source was not detected. 
However, the detection threshold did not reach the low flux level at which the source was detected 
in 1993 (large off-axis angles, short exposures) and no additional information about variability 
can be inferred. The source is probably also identical to the ASCA source AX\,J0054.8$-$7244 
\citep{2003PASJ...55..161Y} which was detected in Nov. 1998 at a flux level of 4.9\ergcm{-13}, 
similar to the low intensity state in May 1993.
\xna\ is a newly discovered X-ray pulsar in the SMC with a period of 701 s. 
While it was the fainter of the two pulsars during the XMM-observation, it was the 
slightly brighter ($\sim$20\%) one during 
the 1993 PSPC observation, the only time when it was detected by ROSAT. Assuming 
the EPIC spectral parameters shows that it was brighter by 40\% during December
2000 compared to May 1993. Eight non-detections by ROSAT and also by ASCA suggest that \xna\
is overall fainter than \xsa, falling below the detection thresholds of the ROSAT and 
ASCA observations (which were, however, close or even above the detection flux) most of 
the time. Both pulsars seem to belong to the class of 
Be/X-ray binaries with long pulse period and moderate intensity variations.

Spectral analysis of both pulsars shows that a power-law with photo-electric absorption does
not reproduce the EPIC spectra properly. Including soft emission components or an exponential 
cut-off yields acceptable fits, but the statistical quality does not allow us to decide between
these models. The major difference between the cut-off model and the two component models
is the resulting absorption column density, which is consistent with the Galactic foreground
value of 6\hcm{20} for the cut-off model. However, the reddening inferred from the 
suggested optical counterparts is incompatible with such low absorption and favours the soft 
component models. On the other hand the absorption is very high for the thermal plasma model
which leads to an implausibly high intrinsic source luminosity in particular for \xna.
A large fraction of the column density may be local to the source and the different 
emission components might suffer different amounts of absorption. 
This could indicate that the spectrum at lower X-ray energies is more complex as it is seen
from other HMXBs in the Magellanic Clouds. E.g. the pulse phase averaged spectrum of
EXO\,053109-6609.2 \citep{2003A&A...406..471H} shows power-law components attenuated by 
different column densities.
A similar behaviour of the two pulsars presented here is expected if the absorption changes 
with pulse phase as it is indicated by the pulse profiles.

With the new discovery of the 701 s pulsar \xna\ the number of pulsars in the SMC has grown 
to 45 \citep{2004astroph0402053C} and future X-ray observations of the SMC promise to find more.


\begin{acknowledgements}
The XMM-Newton project is supported by the Bundesministerium f\"ur Bildung und
For\-schung / Deutsches Zentrum f\"ur Luft- und Raumfahrt (BMBF / DLR), the
Max-Planck-Gesellschaft and the Heidenhain-Stif\-tung. 
\end{acknowledgements}

\bibliographystyle{apj}
\bibliography{mcs,ins,general,myrefereed,myunrefereed,mytechnical}

\end{document}